\documentclass[aps,prl,twocolumn,showpacs,letterpaper,amsfonts,amssymb]{revtex4}
\usepackage{graphicx,rotating,amsmath,color,dsfont}
\newcommand{\kv}{\mathbf{k}} \newcommand{\qv}{\mathbf{q}}
 \newcommand{\be}{\begin{equation}}
\newcommand{\ee}{\end{equation}} \newcommand{\bea}{\begin{eqnarray}}
\newcommand{\eea}{\end{eqnarray}} \newcommand{\up}{\uparrow}
\newcommand{\down}{\downarrow} \newcommand{\bwt}{\begin{widetext}}
\newcommand{\ewt}{\end{widetext}} \newcommand{\ham}{\mathcal{H}}
\newcommand{\unity}{\mathds{1}} 
 
\newcommand{\bsb}{\begin{subarray}} \newcommand{\esb}{\end{subarray}}
\newcommand{\largem}{\!\!}  \newcommand{\mb}[1]{\mathbf{#1}}
\newcommand{\ov}{\bar}
\newcommand{\vecv}[2]{ \left(\largem
 \begin{tabular}{c}
  $#1$ \\ $#2$
  \end{tabular}
  \largem \right) }

\newcommand{\vech}[2]{ \left(\largem
 \begin{tabular}{c}
  $#1$ \: $#2$
  \end{tabular}
  \largem \right) }

\newcommand{\mat}[4]{ \left( \largem
 \begin{tabular}{cc}
  $#1$ & $#2$ \\ $#3$ & $#4$
  \end{tabular}
  \largem \right) }

\begin{document}
\title{Polaron formation in the presence of Rashba spin-orbit
coupling: implications for spintronics}

\author{Lucian Covaci and Mona Berciu} \affiliation{ Department of
Physics and Astronomy, University of British Columbia, Vancouver, BC,
Canada, V6T~1Z1}

\begin{abstract}
We study the effects of the Rashba spin-orbit coupling on the
polaron formation, using a suitable generalization of the Momentum
Average approximation. Contrary to previous investigations 
of this problem, we find that  the spin-orbit interaction
decreases the effective electron-phonon coupling. It is thus
possible to lower the effective mass of the polaron by increasing the
spin-orbit coupling. We also show that when the spin-orbit coupling is
large as compared to the phonon energy, the polaron retains only one of
the spin polarized bands in its coherent spectrum. This has major
implications for the propagation of spin-polarized currents in such
materials, and thus for spintronic applications.
 \end{abstract}

 \date{\today} \maketitle

In the field of spintronics there is a continued effort to develop
means of efficiently manipulating the spin of electrons
\cite{spintronics1}. One widely studied approach is to use spin-orbit
coupling (SO) to lift the spin degeneracy, for example the Rashba SO
coupling \cite{rashba} that arises in a confined system if the quantum
well lacks inversion symmetry. Experimentally, the Rashba effect has
been seen in many systems, {\em e.g.}  semiconductor heterostructures
like GaAs and InAs, surface states of metals like Au(111)\cite{sogold}
and surface alloys like Bi/Ag \cite{sobiag} or Pb/Ag \cite{sopbag}.

Confined two-dimensional (2D) systems may also couple strongly to
optical phonons of the substrate. Tuning of these electron-phonon
(el-ph) interactions was shown to be experimentally viable in organic
single-crystal transistors \cite{hulea}. Strong el-ph coupling is
interesting because it leads to polaron creation, whereas the coherent
quasiparticle is an electron surrounded by a phonon cloud. The polaron
dispersion and effective mass can be significantly renormalized from
those of the bare band electron \cite{feshke}.

An interesting question is whether the el-ph and the SO coupling can
be used in conjuction to tailor differently the properties of the two
bands with different spin. The interplay of the Rashba SO and the
el-ph interactions has been considered previously for continuous
systems with parabolic bands and weak el-ph coupling, using the
self-consistent Born approximation \cite{marsiglio1}. The main
conclusion was that SO enhances the effective el-ph coupling due to a
topological modification of the Fermi surface, namely an effective
reduced dimensionality of the low energy electronic density of
states. In this Letter, we investigate this problem using a suitable
generalization of the Momentum Average (MA) approximation \cite{ma1},
which allows us to investigate non-parabolic bands (tight-binding
models) for any el-ph coupling strength. This is because this method
is accurate for any coupling strength, being exact in the asymptotic
limits of both very weak and very strong coupling. We demonstrate that
the conclusions of Refs. \onlinecite{marsiglio1} do not apply in the
low density limit for a tight-binding dispersion. In fact, the exact
opposite holds, namely the effective el-ph coupling is suppressed by
SO coupling. We also calculate spin-dependent spectral functions and
show that in a certain regime, the coherent polaron band is dominated
by contributions from only one SO electronic band ( the ``-''
band). This has major implications for spin dependent transport
through such materials, allowing for the possibility to manipulate the
effective mass and the spin polarization of quasiparticles by tuning
the el-ph and SO interactions.

We consider a single electron on a two dimensional square lattice with
SO coupling, which also interacts with optical 
phonons of energy $\Omega$ ($\hbar=1$) through a local
Holstein-type interaction~\cite{holstein}. The
 Hamiltonian can be written in terms of $\kv$-space spinors as: \bea
\label{eq:Hamilt}\nonumber
&& \ham=\sum_{\kv}
\vech{c_{\kv\up}^\dagger}{c_{\kv\down}^\dagger}\mat{\epsilon_\kv}{\phi_\kv}
	 {\phi_\kv^\ast}{\epsilon_\kv}\vecv{c_{\kv\up}}{c_{\kv\down}} 
+ \Omega \sum_{\qv} b_\qv^\dagger b_\qv \\ \nonumber && +
\frac{g}{\sqrt{N}} \sum_{\kv,\qv}
\vech{c_{\kv-\qv\up}^\dagger}{c_{\kv-\qv
\down}^\dagger}\mat{b_\qv^\dagger+b_{-\qv}}{0}{0}{b_\qv^\dagger+b_{-\qv}}
\vecv{c_{\kv\up}}{c_{\kv\down}} \eea 
where $\epsilon_\kv=-2t[ \cos(k_xa) + \cos(k_ya)]$ is the 2D
free-electron dispersion for nearest-neighbor hopping and
$\phi_\kv=2V_s[i\: \sin(k_xa) + \sin(k_ya)]$ describes the Rashba SO
coupling. Different dispersions and/or SO couplings can be studied
similarly. As usual, $c^\dagger_{\kv,\sigma}$ is the creation operator
for an electron with momentum $\kv$ and spin $\sigma$, while
$b^\dagger_{\qv}$ creates a phonon of momentum $\qv$.  Both electron
spin channels interact with the phonons through the local lattice
displacement, with an el-ph coupling constant $g$. The lattice has a
total of $N$ sites and periodic boundary conditions, and in the end we
let $N\rightarrow \infty$. Momentum sums are over the Brillouin zone (BZ).

The $2 \times 2$ Green's functions for the non-interacting system (no
el-ph coupling) are defined as: \bea \ov{G_0}(\kv,\omega)&=&\langle 0 |
\vecv{c_{\kv\up}}{c_{\kv\down}}\hat{G}_0(\omega)
\vech{c_{\kv\up}^\dagger}{c_{\kv\down}^\dagger}| 0 \rangle \nonumber \\
&=& \mat{G_{0+}(\kv, \omega)}{e^{i\xi_\kv}G_{0-}(\kv,
\omega)}{e^{-i\xi_\kv}G_{0-}(\kv, \omega)}{G_{0+}(\kv, \omega) } \eea
where $\hat{G}_0(\omega)=[\omega-\ham_0+i\eta]^{-1}$ is the resolvent
corresponding to ${\cal H}_0 = {\cal H}|_{g=0}$. The symmetric and
anti-symmetric Green's functions are:
$$ G_{0\pm}(\kv,
\omega)\!=\!\frac{1}{2}\!\left(\frac{1}{\omega+i\eta-\epsilon_k+|\phi_k|}
\pm \frac{1}{\omega+i\eta-\epsilon_k-|\phi_k|}\right)
$$ and the phase factor is $\xi_\kv=\phi_\kv / |\phi_\kv|$. To avoid
confusion with scalars, all $2\times2$ matrices will be identified by
a bar, hence the $\ov{G}_0(\kv,\omega)$ notation.

The full Green's function, defined as usually: \bea
\ov{G}(\kv,\omega)=\langle 0|\vecv{c_{\kv\up}}{c_{\kv\down}}
\hat{G}(\omega)
\vech{c^\dagger_{\kv\up}}{c^\dagger_{\kv\down}}|0\rangle , \eea can in 
principle be calculated exactly by applying the equation of motion
method. The full Green's function can be related to the
non-interacting one by using the Dyson equation,
$\hat{G}(\omega)=\hat{G}_0(\omega)+\hat{G}(\omega)V \hat{G}_0(\omega)$
- here $V={\cal H}-{\cal H}_0$ is the el-ph interaction. As shown
previously for the Holstein Hamiltonian~\cite{ma1,ma2,ma3}, the
repeated use of the Dyson identity generates a system of equations
involving generalized Green's functions with various numbers of
phonons. In the presence of SO interactions, the equations of
motion are similar, except they are in terms of $2
\times 2$ Green's functions. As a result, we can implement the MA
approximations in the same way, and all conclusions regarding accuracy
for all el-ph coupling strengths, sum rules obeyed exactly by the
spectral weight, etc. remain true. A somewhat analogous procedure was  used 
to  compute the  
$2 \times 2$ Green's function describing rippled graphene
\cite{covaci_graphene}, however there the matrices are related to
different sublattices, not to different spin projections.

For all levels of the MA approximation, the  Green's function can be
written in the standard form: \bea
\label{fullG}
 \ov{G}(\kv,\omega)=[\ov{G_0}(\kv,\omega)^{-1}-\ov{\Sigma}(\kv,\omega)]^{-1},
 \eea 
where the self-energy $\ov{\Sigma}(\kv,\omega)$ has different
 expressions depending on the level of MA approximation used. For the
 simplest, least accurate MA$^{(0)}$ level~\cite{ma1,ma2}, the self-energy has no
 $\kv$ dependence and is given by an infinite 
 continued fraction $\ov{\Sigma}_{MA^{(0)}}(\omega)=g^2
 \ov{A_1}(\omega)$, defined by
 \bea \label{cf}
 \ov{A}_n(\omega)=n\ov{g}_0(\omega-n\Omega)[\unity
 -g^2\ov{g}_0(\omega-n\Omega) \ov{A}_{n+1}(\omega) ]^{-1},  \eea
where
\bea \ov{g_0}(\omega)= {1\over N} \sum_{\kv
} \ov{G_0}(\kv,\omega)=\mat{g_{0+}(\omega)}{0}{0}{g_{0+}(\omega)}
\eea is the momentum average of the noninteracting Green's
 function over the BZ. Note that because the off-diagonal part 
is anti-symmetric, its average over the Brillouin zone
 vanishes, thus $\ov{g_0}(\omega)$ and all $\ov{A}_n(\omega)$ matrices
 are diagonal.

As discussed extensively in Refs. \onlinecite{ma2} and
\onlinecite{ma3},  MA$^{(0)}$ is 
accurate for ground 
 state properties but it fails to properly predict the polaron+one
 phonon continuum. As a result, it overestimates the polaron
bandwith. This problem is fixed at the MA$^{(1)}$ level, where a
 phonon is allowed to appear away from the polaron cloud. For the Holstein
 model (and by extension, in the presence of SO coupling) both these
 approximations predict $\kv$-independent self-energies. Here,  they
 are diagonal as well, {\em i.e.} phonon 
 emission/absorption is not allowed to scatter the electron between
 the two spin-polarized bands. These, of course, are approximations,
 although it is worth pointing that all non-crossed diagrams give
 $\kv$-independent and diagonal contributions to the self-energy. This
 is why the self-consistent Born approximation prediction for the
 self-energy is also $\kv$-independent and
 diagonal~\cite{marsiglio1}. 

In MA, however,  the effect of non-crossed diagrams is included for
MA$^{(2)}$ and higher levels (in variational terms, these allow two or
more phonons to appear away from the main polaron cloud and the order
of their creation/annihilation now becomes relevant). We therefore
report  MA$^{(2)}$ results here. Following a similar procedure to that
described in detail in Ref.~\cite{ma3}, the
MA$^{(2)}$ self-energy is found as:
\bea
\label{selfengma2}
\ov{\Sigma}_{MA^{(2)}}(\kv,\omega)=\ov{x}(0),  \eea 
given by the solution of the
system of coupled equations for the   unknown $2\times2$ matrices $\ov{x}(i)$: \bea
\label{system}
\sum_j \ov{M}_{i,j}(\kv,\omega) \ov{x}(j) = e^{i\kv
\mathbf{R}_i}g^2\ov{G}_0(-i,\tilde{\tilde{\omega}}). \eea The sum is
over lattice sites $i=(i_x,i_y)$ located at
$\mathbf{R}_i=i_xa \hat{x} + i_y a \hat{y}$. The
$2 \times 2$ matrices $\ov{M}_{i,j}(\kv,\omega)$ are:
\bea
\label{sysone} \ov{M}_{00} = \unity- g^2
\ov{g}_0(\tilde{\omega})\ov{g}_0(\tilde{\tilde{\omega}})\left(2
\ov{a}_{31}^{-1} - \ov{a}_{21}^{-1}\right), &&\\
\label{systwo} \ov{M}_{i0} = - g^2 \ov{g}_0(\tilde{\omega})e^{i\mb{k}\cdot {\mb
 R}_i}\ov{G}_0(-i, \tilde{\tilde{\omega}}) \left(2 \ov{a}_{31}^{-1} -
\ov{a}_{21}^{-1}\right) && \eea for $i\ne 0$, and for both $i, j\ne 0$:
\bea \nonumber && \ov{M}_{ij} = \ov{a}_{21} \delta_{i,j} \unity - g^2
e^{i\mb{k}\cdot {\mb R}_i}\ov{G}_0(j,\tilde{\omega})\\
\label{systhree} && \times \left[(\ov{A}_2-\ov{A}_1)\delta_{i,-j} +
 \ov{G}_0(-i-j,\tilde{\tilde{\omega}}) \ov{a}_{21}^{-1}\right].  \eea
Here we defined $\ov{a}_{ij}= \unity -
g^2\ov{g}_0(\tilde{\omega})(\ov{A}_i-\ov{A}_j) $ where
$\ov{A}_1\equiv\ov{A}_1(\omega-2\Omega)$,
$\ov{A}_2\equiv\ov{A}_2(\omega-\Omega)$,
$\ov{A}_3\equiv\ov{A}_3(\omega)$ are continuos fractions defined by
Eq. (\ref{cf}), and
$\tilde{\omega}=\omega-2\Omega-g^2\ov{A}_1|_{(1,1)}$,
$\tilde{\tilde{\omega}}=\omega-g^2\ov{g}_0(\tilde{\omega})|_{(1,1)}(\ov{a}_{21}^{-1})|_{(1,1)}$
(since the $\ov{A}, \ov{a}, \ov{g}_0$ matrices are proportional to
$\unity$, the (2,2) diagonal matrix element can be used just as well
in the definitions of $\tilde{\omega}, \tilde{\tilde{\omega}}$).
Finally, the real space Green's functions appearing in the
inhomogeneous terms are given, as usual, by: \bea
\label{Gi} 
\ov{G}_0(i,\omega)=\frac{1}{N}\sum_{\kv} e^{i\kv\mathbf{R}_i}
\ov{G}_0(\kv,\omega).  \eea It is important to note that for $i \ne
0$, $\ov{G}_0(i,\omega)$ acquires off-diagonal components, which lead
to the off-diagonal contributions in
$\ov{\Sigma}_{MA^{(2)}}(\kv,\omega)$. Because below the free-electron
continuum the $\ov{G}_0(i,\omega)$ decrease exponentially as
$|\mathbf{R}_i|$ increases, the system in Eq. ~(\ref{system}) can be
truncated at a small $|i|$. We truncate at $|\mathbf{R}_i|
\simeq 10 a$, such 
that the relative error of the spectral function is less that $ 10^{-3}$.

\begin{figure}[t]
\centering \includegraphics[angle=-90,width=1.0\columnwidth]{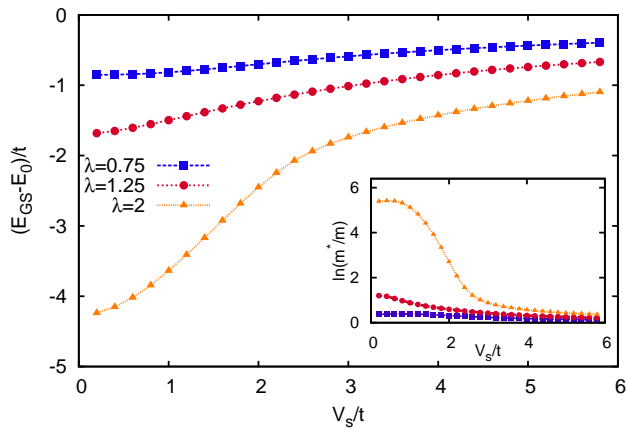}
\caption{(color online) The ground state energy as a function of
Rashba SO coupling for three values of the el-ph coupling. The inset
shows the effective mass on a logarithmic scale as a function of Rashba
SO coupling.}
\label{fig:1}
\end{figure}

Once the self-energy is known, the full Green's function can be
calculated and in turn provide accurate estimates for spectral
weights, the ground state energy, the effective mass etc. In the
non-interacting case ($g=0$) the ground state consists of four
degenerate points in $\kv$-space, $(\pm k_{min},\pm k_{min})$, where
$k_{min}a=\arctan[V_s/(\sqrt{2}t)]$, with the ground-state energy
$E_0=-4t\cos(k_{min}a)-V_s\sqrt{8}|\sin(k_{min}a)|$.  On the other hand,
in the absence of SO coupling ($V_s=0$), as the effective el-ph
coupling $\lambda=g^2/(4t\Omega)$ is turned on, there is a crossover
from a light, large-polaron to a very heavy, small-polaron at $\lambda
\sim 1$.  In Fig.~\ref{fig:1} we show the ground state energy
measured from $E_0$ for weak, medium and strong effective el-ph
couplings, as a function of the Rashba SO coupling.  For large SO
coupling, the renormalization of the energy and effective mass (shown
in the inset) is strongly suppressed, indicating large-polaron
behavior even when $\lambda=2$. This contradicts reported results for
a parabolic band dispersion~\cite{marsiglio1}, which are
based on the fact that the density of states at the band edge has a
square root singularity, because in a continuum model the locus of
momenta defining the ground state is a circle of radius $k_{min}$.
This is not generically true for a tight binding dispersion, where
the Van Hove singularity is shifted from the 4-point degenerate ground
state to higher energies.  As expected, though, the results for a parabolic
dispersion can be recovered by our tight-binding model in the regime
where both $\lambda$ and 
$V_s/t$ are very small. Indeed, for $\lambda=0.75$, the effective mass
increases slightly with $V_s$ 
at small $V_s$, before decreasing at larger $V_s$ values. Such 
behavior is more apparent as $\lambda \rightarrow 0$ \cite{unpubl}.

Our results can be understood by noting that the free-electron
bandwidth increases with increasing $V_s$. This results in an effective el-ph
coupling, which compares the polaron binding energy to this
renormalized bandwidth, that
effectively decreases. As a result, away from the limit where both
$\lambda$ and $V_s/t$ are very small, an increase in the SO coupling
leads to a drop in the effective mass, making it possible to tune
the mass of the polaron between  heavy and light. Light polarons are thus
found for either small 
$\lambda$ irrespective of $V_s/t$,  or at large $\lambda$ and large enough 
$V_s/t$. As we show next, however, their nature and spin-character may
be very different.

The MA$^{(2)}$ approximation is quantitatively accurate not only for
the ground state, but also for high energy states, as it satisfies
exactly the first 8 spectral weight sum rules and with good accuracy
all other ones. It is thus possible to have an accurate depiction of
the high energy states by calculating the spectral function. Because
we expect to have spin polarized bands it is necessary to calculate a
spin dependent spectral function: \bea
\label{spectral}
\vec{A}(\kv,\omega)=-\frac{1}{\pi}Im(Tr[\vec{\sigma}\ov{G}(\kv,\omega)]),
\eea where $\vec{\sigma}$ are the Pauli matrices. The direction of
$\vec{A}(\kv,\omega)$ gives the direction of the expectation value of
the spin, while its magnitude gives the density of states with
momentum $\kv$. We know that in the non-interacting case as we go
around the $\Gamma$-point, eigenstates have spin perpendicular to
their momentum direction and rotating clockwise for one band and
anti-clockwise for the other. For the coupled system we observe
similar spin eigenstates and thus choose to plot the following
quantity: \bea
\label{Atilde}
\tilde{A}(\kv,\omega) = [\vec{u}_k \times \vec{A}(\kv,\omega)] \cdot
\vec{u}_z, \eea where $\vec{u}_k$ and $\vec{u}_z$ are unit vectors
parallel to $\kv$, respectively $z$-axis. The two spin polarized bands
will now correspond to opposite signs of $\tilde{A}(\kv,\omega)$.

\begin{figure}[t]
\centering \includegraphics[angle=-90,
width=\columnwidth]{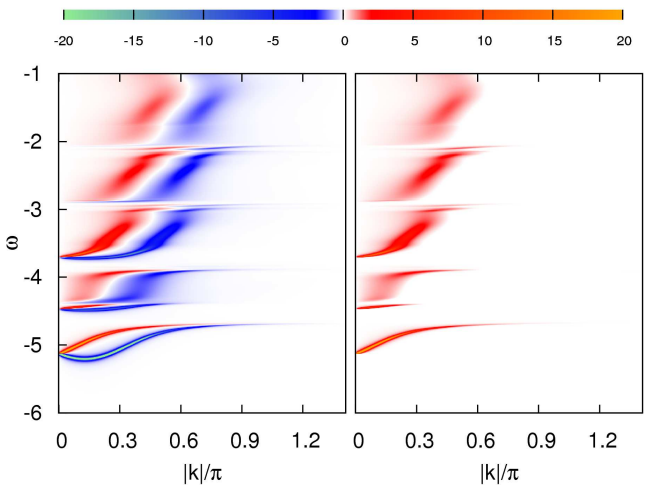}
\caption{(color online) Spin dependent spectral function
$\tilde{A}(\kv,\omega)$, for $\lambda=1$, $\Omega=0.8t$ and
$V_s=0.4t$. The right panel shows only the "+" band for clarity.}
\label{fig:2}
\end{figure}

Since the polaron bandwidth cannot exceed $\Omega$, we expect the
character of the polaron band to depend on the relation between $\Omega$
and the energy difference between the spin-split
electron bands. In order to exemplify this we plot
$\tilde{A}(\kv,\omega)$ in Fig.~\ref{fig:2} for $\lambda=1$,
$\Omega=0.8t$ and $V_s=0.4t$. The spectral function is shown along the
$(0,0)$-$(\pi,\pi)$ line in order to intersect the ground state at
$(k_{min},k_{min})$. In this case $\Omega / (E_0-4t) > 1$ and we see
two coherent polaron bands corresponding to the two spin
polarizations, with similar quasiparticle weights. We conclude that
when $\Omega$ is larger than the SO splitting, the polaronic
quasiparticles are rather similar to the non-interacting electrons, except for
 the renormalized mass  and supressed quasiparticle weight. Of
course, the spectrum above 
$E_{GS}+\Omega$ becomes incoherent due to el-ph scaterring.

\begin{figure}[b]
\centering \includegraphics[angle=-90,
width=\columnwidth]{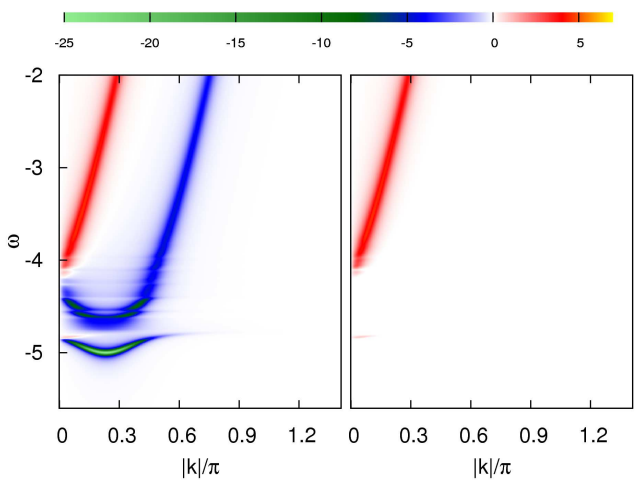}
\caption{(color online) Spin dependent spectral function
$\tilde{A}(\kv,\omega)$, for $\lambda=1$, $\Omega=0.2t$ and
$V_s=0.8t$. The right-hand side panel shows only the "+" band for
clarity.}
\label{fig:3}
\end{figure}

In Fig.~\ref{fig:3} we plot $\tilde{A}(\kv,\omega)$ for
$\Omega=0.2t$, $\lambda=1.0$ and $V_s=0.8t$. Now $\Omega / (E_0-4t) <
1$ and because the  polaron bandwidth cannot exceed
$\Omega$, it is dominated by the ``-'' band. There is a
large difference between the quasiparticle weights of the two
coherent polaron bands (note the different positive and negative scales for the
the contour
plot), and the effective mass of the dominant ``-'' band is much
smaller than that of the low-weight ``+'' band. Higher energy states have
small  weights and are highly incoherent, {\em i.e.} short
lived. 

Consider now  injection of a current in such
a system. Whereas in a regime like that depicted in Fig.~\ref{fig:2} we
expect the spin to precess between the two coherent polaronic bands, like it
does for non-interacting 
electrons~\cite{spin_prec1, spin_prec2}, in a regime like in
Fig.~\ref{fig:3} only the spin-component parallel to the ``-'' band can be
efficiently transmitted through the system, which therefore
acts as an intrinsic ``spin-polarizer''.

This becomes more and more apparent as one moves further into the
asymptotic limit where both the SO and the el-ph couplings are large
compared to $\Omega$, {\em i.e.} $ (E_0-4t) \gg \Omega$ and $\lambda
\gg 1$.  In this limit the ``-'' band in the coherent polaron spectrum
becomes lighter and has a higher quasiparticle weight, whereas the
``+'' band essentially vanishes from the coherent spectrum (its
quasiparticle weight is extremely low and its effective mass is
extremely large). The resulting light polaron is thus very different
from the one we find in the small $\lambda$ regime, which has both
bands in the coherent spectrum with roughly equal quasiparticle
weights and effective mass. 

This demonstrates that an interplay between SO and el-ph couplings
allows indeed for different tailoring of the properties of the two
spin-polarized bands, whereas one is well described by a
long-lived, light quasiparticle while the other becomes highly
incoherent. This will naturally lead to very different conductivities
for the two spin polarizations, making such a material ideal
as a source and/or detector of  spin-polarized currents -- these are 
important components needed for many spintronics applications. These conclusions
are based on the use of the MA approximation which is known to be
highly accurate for all el-ph couplings. Moreover, at the MA$^{(2)}$
level we use here, it results in  a non-diagonal, $\kv$-dependent
self-energy, therefore our results  properly include phonon-mediated
scattering between the two electronic bands. 

Acknowledgments: This work was supported by CIFAR Nanoelectronics and
NSERC. Discussions with Frank Marsiglio are gratefully acknowledged.

\bibliographystyle{revtex}

\end{document}